# The Excess Heat Capacity in Glass-forming Liquid Systems Containing Molecules

H. B. Ke, P. Wen[*], W. H. Wang

Institute of Physics, Chinese Academy of Sciences, Beijing 100190, P. R. China

## Abstract

The excess heat capacity at glass transition temperature in two types of glass-forming systems of $[x\text{NaNO}_3 \cdot (1-x)\text{KNO}_3]_{60}[\text{Ca(NO}_3)_2]_{40}$ ($0 \leq x \leq 1$) and $\text{Ca(NO}_3)_2 \cdot y\text{H}_2\text{O}$ ($4 \leq y \leq 13$) is studied. In the former system, with the replacement of $K^+$ cation with $Na^+$ cation, the excess heat capacity is almost invariable around 65.1 $J \cdot mol^{-1} \cdot K^{-1}$, while the excess increases by 38.9 $J \cdot mol^{-1} \cdot K^{-1}$ with the increasing per molar $H_2O$ content in latter system. A quantitative description of the excess heat capacity is built up with the consideration of atomic and molecular translational motion in the glass-forming systems. This finding might offer further understanding to the glass transition.

## I. Introduction

The excess of the specific heat of glass-forming liquids relative to rigid glass or crystalline materials are of particular importance to understand the mysterious glass transition [1-8]. Due to the nonzero $\Delta C_P$ existing widely in glass-forming liquids with few notable exceptions like fused quartz [1, 2], a nonzero temperature $T_K$ called as Kauzamann temperature, always exist to satisfy the equation, $\Delta S_f = \int_{T_K}^{T_m} \frac{\Delta C_P}{T} dT$, where $\Delta S_f$ and $T_m$ are the fusion entropy and the melting point [1]. The Kauzamann's paradox implies that the glass transition is not only a kinetic process, but also an underlying thermodynamic phase transition. It is consistent with the empirical relation $T_K \approx T_0$ holding for many glass-forming liquids (GFLs) [9, 10]. $T_0$ is Vogel-Fulcher temperature, where the structural relaxation time or viscosity of GFLs is instable hypothetically. Correspondingly, the $\Delta C_P$ has always been described with the dynamic behaviors of GFLs in popular models [1, 11-15] to glass transition. According to the entropy model [15], the $\Delta C_P$ is given as $dS_{Conf}/d\ln T$, where $S_{Conf}$ is the configurational entropy. Unfortunately, in theory it is too difficult to separate the entropy into vibrational and configuratinal parts [16]. In the free volume theory [14], the $\Delta C_P$ is related to the change of the free volume with temperature. However, the free volume in principle is not measured quantitatively, and the quantitative description of the $\Delta C_P$ based on free volume for GFLs is impossible. To describe the heat capacity $C_P$ of GFLs is still a problem unresolved.

According to statistical physics, heat capacity of any matter in equilibrium in principle has a close relation with its intrinsic motions [17]. A basic characteristic of a liquid is well known [18]: A liquid can not support any shear within the normal experimental time scale. If an atom or a molecule in the liquid state is considered at any moment, it will be interacting with the neighbors, vibrating as though it were an atom or a molecule in the solid state. At the next moment with certain gap in time to the previous moment, the atom/molecule may already move away from its previous site. Obviously, the processed involve atomic/molecular translational motion, and need overcome an energy barrier $\Delta E$, and are activated thermally. The average $\Delta E$ in general at constant pressure depends on temperature, and the structural relaxation time in a GFL following the general Arrehnius form can obey the Vogel-Fulcher equation like: $\tau = \tau_0 \exp\left(\frac{\Delta E}{k_B T}\right) = \tau_0 \exp\left(\frac{D}{T-T_0}\right)$, where $k_B$ is Boltzmann's constant and $D$ is a constant. It is consistent with the current elastic model [19] where the barrier for a flow event must be considerably larger than the thermal energy as temperature is approaching to the glass transition temperature. Then, a glass is formed as $\tau$ is so large that the translational motions can not be observed within the experimental time scale. In other words, within the experimental time scale the translational motions are available in GFLs, but the corresponding glasses. Accordingly, the $\Delta C_P$ of glass-forming liquids might be related to the translational motions.

In this letter, we manage to build up a correlation between the excess motions in GFLs relative to their glassy state and the $\Delta C_P$ at $T_g$ in two typical $[x\text{NaNO}_3\cdot(1-x)\text{KNO}_3]_{60} \cdot [\text{Ca(NO}_3)_2]_{40}$ ($0 \leq x \leq 1$) and $\text{Ca(NO}_3)_2\cdot y\text{H}_2\text{O}$ ($4 \leq y \leq 13$) GFL systems. Our experimental results confirm the relation.

**II. Experimental**

We chose $[x\text{NaNO}_3\cdot(1-x)\text{KNO}_3]_{60} \cdot [\text{Ca(NO}_3)_2]_{40}$ ($0 \leq x \leq 1$) and $\text{Ca(NO}_3)_2\cdot y\text{H}_2\text{O}$ ($4 \leq y \leq 13$) glass forming systems since they offer a good opportunity to understand the origin of the $\Delta C_P$ for the systems containing atoms and simple molecules. Besides the good glass-forming ability and a clear $\Delta C_P$ at $T_g$, the motions of the compositional units can be well clarified. The materials were formed directly by the melting of mixtures of deionized water, $\text{NaNO}_3$, $\text{KNO}_3$, and $\text{Ca(NO}_3)_2\cdot 4\text{H}_2\text{O}$. Reagent grade $\text{NaNO}_3$, $\text{KNO}_3$, and $\text{Ca(NO}_3)_2\cdot 4\text{H}_2\text{O}$ were obtained from Sigma-Aldrich, which were directly used without further purification. The detail preparation of $6[x\text{NaNO}_3\cdot(1-x)\text{KNO}_3]\cdot 4[\text{Ca(NO}_3)_2]$ materials and samples used in the measurements of differential scanning calorimeter (DSC) has been reported in Ref.[20]. The $\text{Ca(NO}_3)_2\cdot y\text{H}_2\text{O}$ materials with the mass of 5 g were sealed into glasses vials. In order to homogenize the mixture, the vials were hold at 333 K for 15 min. Thermal analysis occurring during both cooling and heating was carried out using a Mettler Toledo DSC1 thermal analyzer. The mass of the samples used for DSC measurements are around 15 mg.

### III. Results and Discussion

Figure 1 shows the DSC curves for the $[x\text{NaNO}_3\cdot(1-x)\text{KNO}_3]_{60}[\text{Ca(NO}_3)_2]_{40}$ ($0 \leq x \leq 1$) materials. All of curves were measured at a heating rate of 20K/min. The measurements followed a cooling from the temperature ($T_g+20$ K) at 20 K/min for each DSC sample. The system can be regarded as a binary mixture. Potassium, sodium and calcium ions have spherical charge distributions and different from nitrate ions of trigonal shape [21]. It has been reported that the glass-forming ability of $[x\text{NaNO}_3\cdot(1-x)\text{KNO}_3]_{60}[\text{Ca(NO}_3)_2]_{40}$ ($0 \leq x \leq 1$) systems become worse with the increase of $x$ [20]. Even so, a clear glass transition with an obvious $\Delta C_P$ and supercooled liquid region still exists for all of compositions. The $T_g$, determined by the usual way (see Fig.1), is found to depend on the composition. With the replacement of the first alkali ionic ($K^+$) by the second alkali ionic $Na^+$, $T_g$ for the system decreases gradually to a minimum as $x$ approaching to 0.5. A similar phenomenon exists in another system [22] and represents the alkali-mixed effect [23]. The fragility index $m$ of GFLs can be given by DSC measurements [24]. The determined values of $T_g$, $\Delta C_P$ at $T_g$ and $m$ for the system are listed in Table I.

The $x$ dependence of the fragility index $m$ and the $\Delta C_P$ of the $[x\text{NaNO}_3(1-x)\text{KNO}_3]_{60}\cdot[\text{Ca(NO}_3)_2]_{40}$ system are displayed in Fig.2. For $[\text{KNO}_3]_{60}[\text{Ca(NO}_3)_2]_{40}$, CKN, with extreme $x = 0$, the value of $m$ is equal to 107. The value of 107 is higher than the value $m = 93$ reported by Bohmer *et al.*,[25], and less than the value 125 obtained from the temperature dependence of viscosity close to $T_g$ [26]. The value of $m$ for the system depends on the composition, and the tendency of $m$ with $x$ is similar to that of $T_g$. As $x$ tends to 0.5, $m$ approaches to a minimum. In contrast, the $\Delta C_p$ is almost independent of the composition and close to a constant value of 65.2 J·K$^{-1}$·mol$^{-1}$. This implies that in this system no relation between the $\Delta C_p$ and the $m$ exists. Therefore, the conventional wisdom that GFLs with large *m* has large $\Delta C_p$ at $T_g$[12, 27], can not be applied in this GFL system. A similar result has been found in oxide glass-forming systems [22].

As a binary system, the carriers of the motions only have two types. One type contains $K^+$, $Na^+$, and $Ca^{2+}$ ions. Another involves $NO_3^-$ ions. The contributions of $K^+$, $Na^+$, and $Ca^{2+}$ ions to the heat capacity are equivalent since the $\Delta C_p$ is invariable as $K^+$ is replaced by $Na^+$ in system. It is appropriate to separate the $\Delta C_p$ into cationic and anionic parts. Defining the $C_P^{ca}$ related to one mole cations and the $C_P^{an}$ for one mole anionic $NO_3^-$, the $\Delta C_p$ in $[x\text{NaNO}_3\cdot(1-x)\text{KNO}_3]_{60}[\text{Ca(NO}_3)_2]_{40}$ system can be expressed as: $\Delta C_P = C_P^{ca} + 1.4 \cdot C_P^{an} = 64.9 JK^{-1}mol^{-1}$. To determine the values of the $C_P^{ca}$ and $C_P^{an}$, another equation containing the $C_P^{ca}$ and $C_P^{an}$ must be provided.

Figure 3 shows the DSC curves for a series of $Ca(NO_3)_2 \cdot yH_2O$ ($4 \leq y \leq 13$) samples. The DSC measurements are carried out at a heating of 10 K/min. The values of $T_g$ and the $\Delta C_p$ at $T_g$ for all of materials are listed in Table I. In this system, $T_g$ decreases obviously with the increasing $H_2O$ content, but the $\Delta C_p$ increases. The values of $T_g$ and $\Delta C_p$ for $Ca(NO_3)_2 \cdot 4H_2O$ are 212 K and 234.3 J·K$^{-1}$·mol$^{-1}$. We note that the values are a little smaller than those reported previously [7]. It is due to the different measurements used with different samples. Importantly, we will show the compositions, especially the $H_2O$ content, have a great effect on the $\Delta C_p$ (see Table I and Fig.4). The large value of $\Delta C_p$ has been explained since its *m* (around 100) is large and its GFL is a typical fragile one [7, 28]. The value of *m*, compared with that of CKN, is still smaller, but $Ca(NO_3)_2 \cdot 4H_2O$ has much larger $\Delta C_p$ than CKN has. It can not be explained well by the conventional wisdom [12, 27]: the more *m* is, the larger $\Delta C_p$ is. Different from the conventional wisdom, we think the key reason is that the particles involving $Ca^+$, $NO_3^-$, and $H_2O$ in one molar $Ca(NO_3)_2 \cdot 4H_2O$ sample is more than those in one molar CKN.

Following the description of the $\Delta C_p$ in $[x\text{NaNO}_3 (1-x)\text{KNO}_3]_{60} \cdot [\text{Ca(NO}_3)_2]_{40}$ system, the $\Delta C_p$ for $Ca(NO_3)_2 \cdot yH_2O$ system is written as $C_P^{ca} + 2 \cdot C_P^{an} + y \cdot C_P^{H_2O}$, where $C_P^{H_2O}$ is related to one molar $H_2O$ molecules. In $Ca(NO_3)_2 \cdot yH_2O$ systems, the only variable is $H_2O$ content. The change $\Delta C_p$ in the system must be proportional to the change of $y$. Moreover, the values of the $C_P^{H_2O}$ and the sum $C_P^{ca} + 2 \cdot C_P^{an}$ can be deduced from the linear relation between the $\Delta C_p$ and $y$. The linear increase of the $\Delta C_p$ with the increase of $y$ in the system is shown in Fig. 4. The slop of the linear relation is 37.9±1.1 J·mol$^{-1}$·K$^{-1}$, and its intercept on vertical axis is 87.6±10 J·mol$^{-1}$·K$^{-1}$. So, the value of $C_P^{H_2O}$ is 37.9 J·mol$^{-1}$·K$^{-1}$, and the sum of $C_P^{ca} + 2 \cdot C_P^{an}$ is equal to 87.6 J·mol$^{-1}$·K$^{-1}$. Combined with the result, $C_P^{ca} + 1.4 \cdot C_P^{an} = 64.9 JK^{-1}mol^{-1}$, the values of the $C_P^{ca}$ and $C_P^{an}$ are determined to be 12.0 and 37.8 J·mol$^{-1}$·K$^{-1}$. Remarkably, it is found that the value of the $C_P^{ca}$, $C_P^{an}$ and $C_P^{H_2O}$ can be written as $\frac{3}{2}R$ (around 12.5 J·mol$^{-1}$·K$^{-1}$), $\frac{9}{2}R$ (around 37.4 J·mol$^{-1}$·K$^{-1}$) and $\frac{9}{2}R$ with a small error, respectively.

The following question is what contributes to the $C_P^{ca}$, $C_P^{an}$ and $C_P^{H_2O}$. It is usual, in theoretical considerations, to ignore the difference between the $C_P$ at constant pressure and $C_V$ at constant volume for a condensed matter; this neglect involves only small errors, and can be remedied [29]. In theory, the $C_V$ can be calculated directly from the motions of particles, whose additive integrals completely define the statistical properties of a closed system [17]. The motions of particles in a liquid involve vibrations, rotations and translations. Clearly, atomic vibrations in the GFLs have nothing with the $C_P^{ca}$. The vibrations for one atom have six degrees of freedom, and contribute $\frac{6}{2}k_B$ to the heat capacity, and the atomic vibration can exist in both of glass and liquid. The remaining atomic motions, contributing to the $C_P^{ca}$, are atomic translational motions. These motions are related to the atomic jumps away from their positions in the supercooled liquid state. In space, each atomic translational motion has three independent directions with the same possibility. So, this motion has three degrees of freedom. The corresponding heat capacity arisen from atomic translation is $\frac{3}{2}R$, and exactly equal to the experimental value of $C_P^{ca}$.

It is appropriate to think that the translational contribution to the heat capacity is not measured in glass region since the structural relaxation of equilibrium liquid can not exist in the normal DSC experimental time. Therefore, $C_P^{ca}$ is determined by the atomic translation.

Compared to atomic motions, the motions of $NO_3^-$ atomic group and $H_2O$ molecule are a little complex, but still clear. $NO_3^-$ has not only vibration, but also rotation and translation. The vibration related to $NO_3^-$ atomic group involves the vibration of the atomic group and atoms inside the atomic group. The former has six degrees of freedom, and the latter has twelve degrees of freedom. So, the $C_P^{an}$ ($\frac{9}{2}R$) does not correspond to the vibration. The rotation of $NO_3^-$, involving the rotation of trigonal plate and the rolling of oxygen around nitrogen inside the atomic group, has three degrees of freedom in totally, and can not contribute to $C_P^{an}$ either. The translation of $NO_3^-$ atomic group has nine degrees of freedom, but three. It is due to that one $NO_3^-$ atomic group has three molecular forms to translate in space. The three molecular forms correspond to its three degrees of vibrational freedom. One $NO_3^-$ atomic group with a given molecular form has three degrees of translational freedom. So, the translation of one molar $NO_3^-$ atomic group contributes $\frac{9}{2}R$ to the heat capacity, and is the origin of the $C_P^{an}$. Same as $NO_3^-$ atomic group, the correlation between $C_P^{H_2O}$ and the translation of $H_2O$ molecule also exists. One $H_2O$ molecule, regarded as a rigid bent, has three degrees of rotational freedom. Its translation in liquid has nine degrees of freedom, and contributes $\frac{9}{2}R$ to the heat capacity. The effect of molecular rotation on its form to translate in liquid state is due to that the particles have strong interactions with their neighbors. Prior to translation, rotation already takes place. It is known that the activation of molecular rotations is easier than that of the translations in condensed state [30, 31].

Our finding offers a clear picture to the glass transition in GFLs with an obvious $\Delta C_P$. Upon cooling (see Fig. 3), across the glass transition the rapid decrease in heat capacity is accompanied by the gradual disappearance of translations in the GFLs within the DSC experimental time scale. So, the glass transition is a pure kinetic process. The process corresponds to the disappearance of translations within the experimental time scale. Since $T_g$ depends on the experimental time scale, in theory $T_g$ can be any temperature in liquid region below $T_m$. The contribution of the translation to heat capacity at constant pressure is independent of temperature. Thus, the value of the $\Delta C_P$ can be predicted to be invariable with temperature. The finding gives also strong evidence to an idea that the excess entropy of GFLs is only related to the translational motions. It is not necessary to assume that the excess entropy has a close relation with the configurational entropy. Even if the experimental time scale is long enough and $T_g$ should tend to lower temperature, no thermodynamic phase transition could take place at lower $T_g$. The frozen of the translation across the glass transition is just a dynamic process on a certain experimental time.

## IV. Conclusion

In summary, the clear correlation of the excess heat capacity with the corresponding translational motion in $[xNaNO_3 \cdot (1-x)KNO_3]_{60}[Ca(NO_3)_2]_{40}$ ($0 \leq x \leq 1$) and $Ca(NO_3)_2 \cdot yH_2O$ ($4 \leq y \leq 13$) glass-forming systems is built up. The results show that within the DSC experimental time scale all of translational motions are available in supercooled liquids, but in glasses. The glass transition in the two glass-forming systems is related to the frozen of translational motions of all carriers in the GFLs.

The work was supported by the Science Foundation of China (Grant Nrs: 51071170 and 50921091) and MOST 973 of China (No. 2007CB613904, 2010CB731603).

# Captions

Figure 1. The DSC curves of $[x\mathrm{NaNO_3}\ (1-x)\mathrm{KNO_3}]_{60}\cdot[\mathrm{Ca(NO_3)_2}]_{40}$ ($0 \leq x \leq 1$) glassy samples measured during a heating at 20 K/min.

Figure 2. The $x$ dependence the $\Delta C_p$ at $T_g$ and the fragility index $m$ for $[x\mathrm{NaNO_3}\ (1-x)\mathrm{KNO_3}]_{60}\cdot[\mathrm{Ca(NO_3)_2}]_{40}$ glass-forming system.

Figure 3. The DSC curves for series of $\mathrm{Ca(NO_3)_2}\cdot y\mathrm{H_2O}$ ($4 \leq y \leq 13$) glassy samples measured during a heating at 10K/min. The dash lines are the curves measured during a cooling at 10K/min.

Figure 4. The linear relation between the $\Delta C_p$ at $T_g$ and $y$ in $\mathrm{Ca(NO_3)_2}\cdot y\mathrm{H_2O}$ system. The slope of the fitting line is 37.8 J·mol$^{-1}$ ·K$^{-1}$, and interact on vertical axis is 87.6 J·mol$^{-1}$ ·K$^{-1}$.

Table I. The compositions, glass transition temperature $T_g$, the fragility index $m$ and the jumps in heat capacity $\Delta C_p$ at $T_g$ for $[x\mathrm{NaNO_3}\cdot(1-x)\mathrm{KNO_3}]_{60}\cdot[\mathrm{Ca(NO_3)_2}]_{40}$ ($0 \leq x \leq 1$) and $\mathrm{Ca(NO_3)_2}\cdot y\mathrm{H_2O}$ ($y = 4\sim13$) glass-forming systems.

Table I. The values of $T_g$, $\Delta C_P$ at $T_g$ and $m$ determined by DSC for the Ca(NO$_3$)$_2$·$y$H$_2$O and [$x$NaNO$_3$·(1-$x$)KNO$_3$]$_{60}$·[Ca(NO$_3$)$_2$]$_{40}$ ( 0 ≤ $x$ ≤ 1) materials.

| Ca(NO$_3$)$_2$·$y$H$_2$O | | | [$x$NaNO$_3$·(1-$x$)KNO$_3$]$_{60}$·[Ca(NO$_3$)$_2$]$_{40}$ | | | |
|---|---|---|---|---|---|---|
| $y$ | $T_g$ (K) | $\Delta C_p$ (J·mol$^{-1}$·K$^{-1}$) | $x$ | $T_g$ (K) | $m$ | $\Delta C_p$(J·mol$^{-1}$·K$^{-1}$) |
| 4 | 212 | 234.3 | 0 | 339 | 105 | 65.5 |
| 5 | 202 | 281.3 | 0.125 | 336 | 91.5 | 63.6 |
| 6 | 194 | 295.5 | 0.25 | 335 | 83.7 | 66.5 |
| 7 | 187 | 358.5 | 0.375 | 335 | 80.5 | 63.8 |
| 8 | 182 | 397.5 | 0.5 | 334 | 66.7 | 64.4 |
| 9 | 178 | 434.3 | 0.625 | 335 | 87.1 | 65.6 |
| 10 | 174 | 477.5 | 0.75 | 335 | 94 | 63.3 |
| 11 | 171 | 504.2 | 0.875 | 337 | 102.5 | 65.4 |
| 12 | 168 | 534.8 | 1 | 341 | 104 | 65.8 |
| 13 | 166 | 570.2 | | | | |

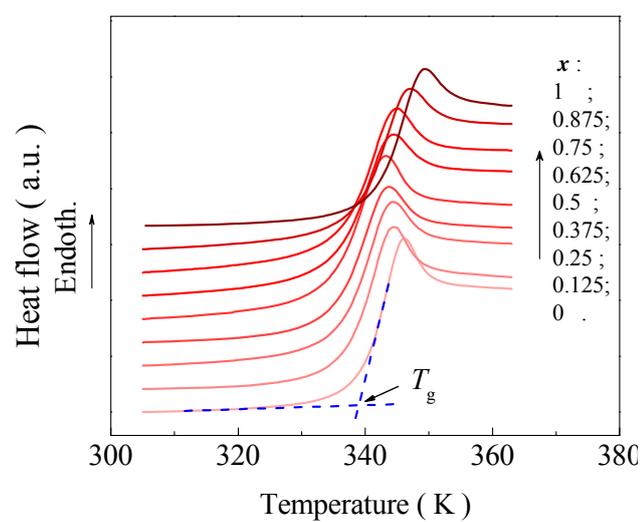

Figure 1. Ke, *et. al.*

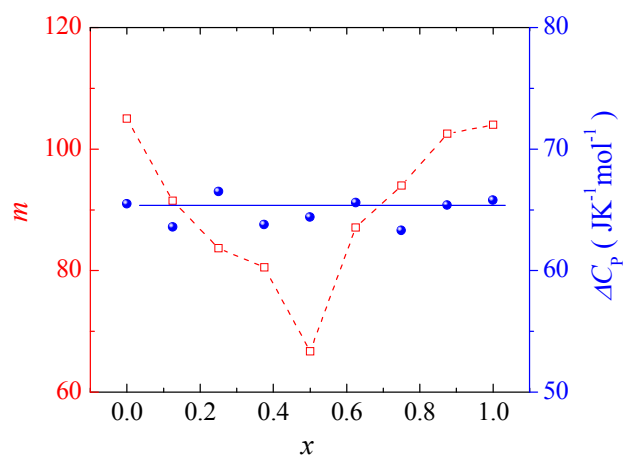

Figure 2. Ke, *et. al*.

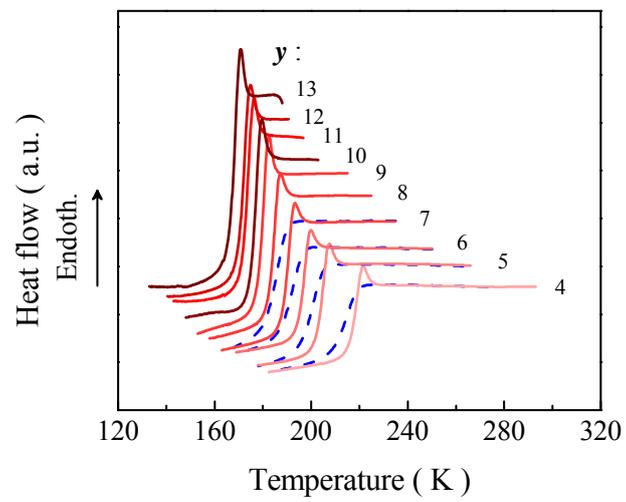

Figure 3. Ke, *et. al*.

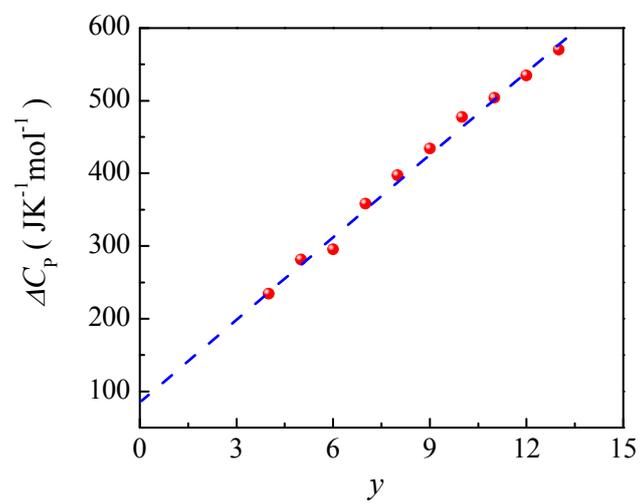

Figure 4. Ke, *et. al*.